# VERIFICATION OF EINSTEIN'S FORMULA FOR GRAVITATIONAL DEFLECTION OF LIGHT USING OBSERVATIONS OF GALACTIC MICROLENSING

**A.N. Alexandrov, V.M. Sliusar, V.I. Zhdanov**

Astronomical observatory, Taras Shevchenko National University of Kyiv, Kyiv, Ukraine

Abstract

*The potential of the gravitational microlensing inside our Galaxy for testing the Einstein's formula for the gravitational light deflection is discussed. For this purpose, the lens mapping is modified by introducing parameter $\varepsilon$, which characterizes the deviation from this formula. An example of such deviation described by a simple power law is analyzed. We formed a sample of 100 microlensing light curves using the data of the Optical Gravitational Lensing Experiment (OGLE) for 2018. The resulting $\varepsilon$ value does not contradict the General Relativity within 1% errors.*

## 1. INTRODUCTION

The General Theory of Relativity (GR), which recently celebrated its centenary, has a strong experimental and observational base, which includes a number of relativistic gravity tests with fairly high accuracy [1,2]. Currently, there is no serious data that could cast doubt on the use of GR for solving astrophysical problems. Nevertheless, it is natural that the emergence of new experimental capabilities stimulates the continuation of such tests.

The Einstein's formula for light deflection angle $\alpha$ in the gravitational field of point mass $M$ is one of the classical predictions of general relativity (see, e.g., [3,4])

$$\alpha(p) = \frac{4GM}{c^2 p}, \qquad (1)$$

where $p$ is the impact parameter. This formula forms the basis of the theory of gravitational lensing and in this sense is confirmed by its achievements [3,4]. An important line of research is connected with to the Galactic gravitational microlensing when a distant source is lensed by a foreground star of our Galaxy [5]. The possibility of using gravitational lensing to test gravitational theories has been discussed elsewhere (e.g., [6-12]), in particular, for comparing GR with modified gravitational theories. To verify the Einstein's formula, it is natural to use the significant dataset on the Galactic microlensing, compiled during the Optical Gravitational Lensing Experiment (OGLE), which started in 1992 [13,14]. Of course, the deflection of light by stars is very small and it has never been directly observed[1] unlike similar effects in the Solar system. However, due to the enormous interstellar distances, the related effect comes into play – the brightness increase of a remote source (e.g., star in the Magellanic Clouds or in the Galactic bulge), when a foreground star passes near the line of sight.

The aim of the present paper is to assess the potential of Galactic gravitational microlensing for testing the relativistic deflection of light. The accuracy of this testing using one microlensing light curve of a remote star is expected to be significantly lower than, for example, by using estimates of the PPN parameter γ in the Solar system. But the existence of an extensive and fairly homogeneous set of photometric data [13, 14] gives us hope for improved accuracy in the future with the accumulation of data. We note that, in comparison with studies of the gravitational effect on electromagnetic radiation in the Solar system, the data on galactic microlensing are completely independent; they are based on a different approach and the other observational material.

In the vast majority of Galactic events, the microlensing mass and the source can be considered as point ones. In this paper we confine ourselves to the microlensing events when the contributions of other objects can be neglected, and the relative motion of the point radiation source and point mass can be considered inertial.

---

[1] See, however, Dong S., et al., 2019, ApJ, 871, 70

To test the formula for gravitational deflection of light, we use the publicly available data of OGLE, namely we use the data of the fourth stage OGLE-IV[2] for 2018.

## 2. BASIC RELATIONS

In the gravitational lensing theory, the key role is played by lens mapping, which relates the angular position **y** of a source on the celestial sphere with the angular position **x** of its image (see, e.g. [3,4]). The normalized lens mapping which corresponds (1) in the case of a point mass lens (Schwarzschild lens) can be written as

$$\mathbf{y} = \mathbf{x}\left(1 - \frac{R_E^2}{|\mathbf{x}|^2}\right), \quad R_E = \left[4GMD_{LS}/\left(c^2 D_S D_L\right)\right]^{1/2}; \qquad (2)$$

where $R_E$ is the angular radius of the Einstein ring, $D_S$ and $D_L$ - distances to the source and to the lens; $D_{LS}$ from the lens to the source; $\mathbf{y}, \mathbf{x}$ may be considered respectively as two-dimensional vectors in the source and lens planes tangent to unit sphere. The coordinate origins are chosen at the center of the point mass $M$ in **x** plane and its projection onto the source plane.

To check formula (1), we will slightly "mess up" (2) by setting

$$\mathbf{y} = \mathbf{x} F(\xi, \varepsilon), \qquad \xi \equiv R_0/r, \quad r = |\mathbf{x}|, \qquad (3)$$

where $R_0$ is an analog of $R_E$ (in angular measure), dimensionless parameter $\varepsilon$ describes a deviation of the right-hand side of (2) from the general relativistic expression $F(\xi, 0) = 1 - \xi^2$. Taking into account the existing gravitational experiments (see, e.g., [1,2]), it is natural to assume that the deviation and parameter $\varepsilon$ are small. In observations of the Galactic microlensing events, the mass of the microlensing star is not known and $R_0$ is a free parameter, which is determined by fitting the light curve. Our task is obtain restrictions on the deviation parameter $\varepsilon$ in case of a concrete function $F(\xi,\varepsilon)$ (and/or maybe a plausible estimate of the deviation, if any).

---
[2] http://ogle.astrouw.edu.pl/ogle4/ews/ews.html

The amplification of a separate image of the source at the point $\mathbf{y}$ is (cf., e.g., [10])

$$K(r,\varepsilon) = \left| F \left[ \frac{d}{dr}(rF) \right] \right|^{-1}, \quad F \equiv F(R_0/r, \varepsilon), \tag{4}$$

where the image position $\mathbf{x}$ must be found from the lens equation (3); $r = |\mathbf{x}|$. For $\varepsilon = 0$ equation (3) is identical to equation (2) which has two solutions for $r \neq 0$ corresponding to two different images. If $F(\xi,\varepsilon)$ is a smooth function of its variables for $r \neq 0$, then in the neighborhood of $\varepsilon = 0$ the number of images does not change, i.e. we have two images $\mathbf{x}_1(\mathbf{y},\varepsilon)$, $\mathbf{x}_2(\mathbf{y},\varepsilon)$. In the case of Galactic microlensing, various images of a remote source are not optically resolved; therefore, the total amplification of all images is required to fit the light curves:

$$K_{tot}(y,\varepsilon) = K(|\mathbf{x}_1(y,\varepsilon)|,\varepsilon) + K(|\mathbf{x}_2(y,\varepsilon)|,\varepsilon). \tag{5}$$

We assume the straight-line uniform motion of the source, then

$$y(t) = Y(t,p,V,t_0) \equiv \sqrt{p^2 + V^2(t-t_0)^2},$$

where $p$ - is the impact parameter, $V$ is the velocity, $t_0$ is a moment of maximum brightness. For magnitudes, we have

$$m(t,\varepsilon,p,V,t_0,C) = -2.5\lg\{K_{tot}(Y(t,p,V,t_0),\varepsilon)\} + C,$$

where $C$ is a constant that is associated with the brightness of the source in the absence of lensing and the choice of standard brightness. Our task is to minimize the function

$$H(\varepsilon,p,V,t_0,C) = \sum_{i=1}^{N} W_i \{m_i + 2.5\lg[K_{tot}(Y(t_i,p,V,t_0),\varepsilon)] - C\}^2, \tag{6}$$

yielding the minimization parameters $p,V,t_0,C,\varepsilon$; here $m_i$ − experimental points on the light curve at moments $t_i$, $W_i$ are the weights, $i = 1,...,N$. We considered two choices of $W_i$: equal weights (uniform measurements) and $W_i = (\Delta m_i)^{-2}$, where $\Delta m_i$ are error estimates according to OGLE database.

An analogous procedure can be performed in terms of measured fluxes, where the weights must be correspondingly recalculated from $W_i$.

## 3. LENS MAPPING MODEL

Now we restrict ourselves to the modification of the formula (1) for the deflection angle as

$$\alpha = \left(\frac{4GM}{c^2 p}\right)^{1+\varepsilon}. \tag{7}$$

Standard considerations yield the function $F(\xi, \varepsilon)$ of the lens mapping (3)

$$F(\xi, \varepsilon) = 1 - \xi^{2+\varepsilon}; \quad \xi \equiv R_0/r, \quad R_0^{2+\varepsilon} = R_E^2 \left(\frac{4GM}{c^2 D_L}\right)^{\varepsilon}. \tag{8}$$

Note that the standard parameterized post-Newtonian (PPN) formalism, which is often used for discussions of the GR tests, cannot be directly applied to (8) as this formula deals with non-analytical expressions. Equation (3) in angular variables corresponding to (7) takes on the form:

$$\mathbf{y} = \mathbf{x}\left[1 - \left(\frac{R_0}{r}\right)^a\right], \quad a = 2 + \varepsilon. \tag{9}$$

This yields

$$|\mathbf{y}| = r\left[1 - \left(\frac{R_0}{r}\right)^a\right]. \tag{10}$$

Simple analysis shows that for small $\varepsilon$ and arbitrary $y = |\mathbf{y}| > 0$ there are two solutions of (10) with respect to $r$: one $r_1 \in (0, R_0)$ and the other $r_2 > R_0$, which goes into $r \approx y$ for large $y$. The solutions can be easily obtained numerically. Correspondingly, for the solutions of equation (9) for $\mathbf{y} \neq 0$ we have

$$\mathbf{x}_1 = -\mathbf{n} r_1, \quad \mathbf{x}_2 = \mathbf{n} r_2, \quad \mathbf{n} = \mathbf{y}/|\mathbf{y}|.$$

The amplification coefficient (4) of a separate image is

$$K(r,\varepsilon) = \left[1-\left(\frac{R_0}{r}\right)^a\right]^{-1}\left[1+(a-1)\left(\frac{R_0}{r}\right)^a\right]^{-1}. \qquad (11)$$

The total coefficient $K_{tot}(y,\varepsilon)$ is the sum (5) of amplifications of two images for the positions $\mathbf{x}_1(y,\varepsilon), \mathbf{x}_2(y,\varepsilon)$ that are the solutions of (9).

We proposed in [12] to use an expansion with respect to $\varepsilon$, which was implemented for the specific form of function $F$ in (3); analytical expressions for $\mathbf{x}_1(y,\varepsilon)$, $\mathbf{x}_2(y,\varepsilon)$ and amplification (11) were derived as the linear approximation in $\varepsilon$. This approach made it possible to explicitly present the corrections to the light curve due to $\varepsilon$. This can be useful when considering various options of $F$.

## 4. LIGHT CURVE PROCESSING

The OGLE data [13,14] are used in a number of observational programs, such as the classification of variable stars, the detection of microlensing events, dwarf novae, search for exoplanets, and the study of Magellanic Clouds. In this work, we performed a visual inspection of the OGLE data for 2018 from the OGLE website. On this basis, a sample of one hundred strong microlensing events was selected of the first 700 events for 2018. The selection criteria were sufficient brightness enhancement at the maximum, sufficient accuracy and the number of points on the microlensing light curve, the absence of obvious signs of a binary system or planetary contribution (see, e.g., [15-17]) and absence of parallax effects [18,19]. The sample is presented in Appendix A.

The processing algorithm for the *j*-th lighcurve ($j = 1,...,M$, $M = 100$ is the sample size) is as follows. During the first stage of the processing, we fitted the lightcurves within the standard Schwarzschild model [1] ($\varepsilon = 0$) yielding zero approximation parameters $p_j^{(0)}, V_j^{(0)}, t_{0j}^{(0)}, C_j^{(0)}$ for $\varepsilon = 0$ on *j*-th lighcurve. At the second stage these parameters were used as seed values for the minimization procedure applied to function (6) to obtain the final residuals

$$RSS_j = \min\{H_j(\varepsilon, p, V, t_0, C)\} \qquad (12)$$

and fitting parameters $\varepsilon_j, p_j, V_j, t_{0j}, C_j$.

We processed the data using three methods described below, which mainly differ in the choice of weights.

(i) In the first method, we used the linear approximation for the solutions of the lens equation (9) according to [12] with magnitudes recalculated to fluxes, and all weights were taken equal (uniform measurements).

The next two methods deal directly with the stellar magnitudes; equation (9) has been solved numerically.

(ii) For every light curve all weights were taken equal. Note that the choice of equal $W_i$ corresponds to unequal weights of case (i).

(iii) The weights in (6) for each light curve were taken as $W_i = (\Delta m_i)^{-2}$, where $\Delta m_i$ are the error estimates according to the OGLE database.

After calculation of $\varepsilon_j$ from $M$ light curves of the sample we get $\langle\varepsilon\rangle = \sum_{j=1}^{M} w_j \varepsilon_j$ and its standard deviation $\sigma_{\langle\varepsilon\rangle}$; the weights $w_j$ has been chosen according to $w_j \propto 1/RSS_j$, $\sum_{j=1}^{M} w_j = 1$. The median of the $\varepsilon$-distribution has been also estimated. Table 1 below shows the statistical estimates by means of three different methods.

Strictly speaking, this choice of $w_j$ is appropriate for the linear least-squares method. We compared this choice with $w_j \propto (\Delta\varepsilon_j)^{-2}$, where $\Delta\varepsilon_j$ is the marginal dispersion estimate obtained on account of $j$-th microlensing light curve by means of the Monte Carlo simulations within method (ii). The results of comparing different $w_j$ choices are in a satisfactory agreement.

The distribution of $\varepsilon_j$ values in the sample obtained by different methods are presented in Fig. 1. Here $p(\varepsilon)$ is a fraction of microlensing events for which the value of $\varepsilon$ falls into the corresponding bin.

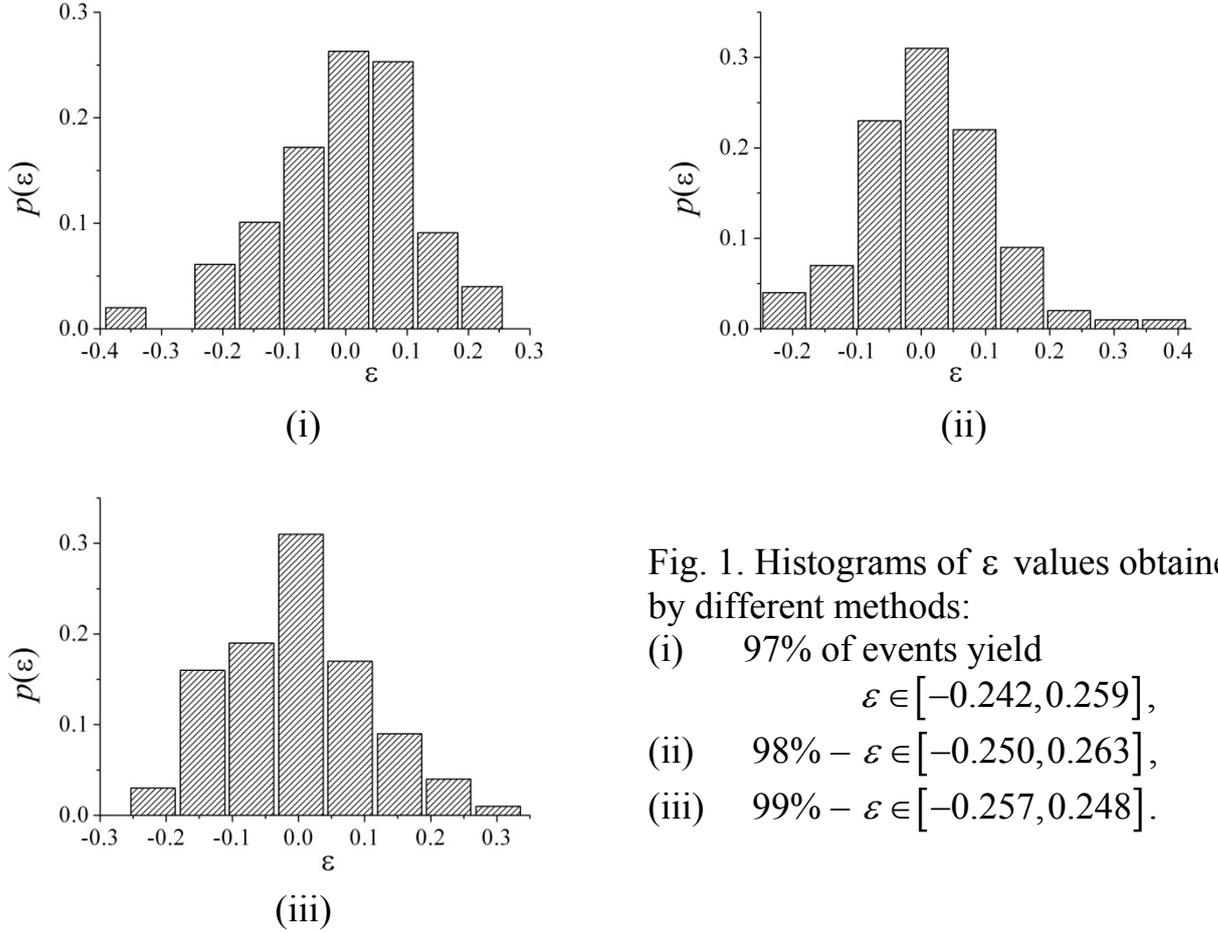

Fig. 1. Histograms of $\varepsilon$ values obtained by different methods:
(i)   97% of events yield
$$\varepsilon \in [-0.242, 0.259],$$
(ii)  98% – $\varepsilon \in [-0.250, 0.263]$,
(iii) 99% – $\varepsilon \in [-0.257, 0.248]$.

Final results are given in Table 1: the values of the median, the average $\langle\varepsilon\rangle$ and the standard deviation $\sigma_{\langle\varepsilon\rangle}$ for distribution of *M* values of $\varepsilon$ obtained by three described methods.

Table 1. The median, mean and standard deviation values for the distribution of $\varepsilon_j$ obtained by three methods.

| Method | median | $\langle\varepsilon\rangle$ | $\sigma_{\langle\varepsilon\rangle}$ |
|---|---|---|---|
| (i) | 0.0053 | 0.0066 | 0.0085 |
| (ii) | 0.0079 | 0.015 | 0.0087 |
| (iii) | 0.0003 | −0.0036 | 0.011 |

We see that the errors in the individual measurements can be relatively large, however, the average $\langle\varepsilon\rangle$ is of the order of $1\sigma$ estimate; i.e., it is statistically insignificant.

## 5. DISCUSSION

The aim of this paper was to assess the possibilities to use the Galactic microlensing light curves for testing the Einstein's formula for the gravitational light deflection.

As can be seen from the Table 1 in all three approaches considered we have the r.m.s. error $\sigma_{\langle\varepsilon\rangle} \sim 10^{-2}$, and average $\langle\varepsilon\rangle$ is approximately of the same order. Therefore, the deviation of the $\varepsilon$ from zero should be considered statistically insignificant. This is an additional confirmation of the Einstein's formula, which demonstrates an alternative method for testing the relativistic gravity.

To discuss the significance of our results, we note that they deal with the observational data that are completely independent of experiments in the Solar system. In this paper we confined ourselves to a concrete and most simple modification (6) of the lens mapping, which involves the power-law dependence upon the impact parameter. However, the use of the microlensing light curves allows us to consider the other forms of the lens mapping.

In the strong microlensing effects in our Galaxy with a remote source for $D_S \gg D_L$, the typical angular distance between the source and the microlens is of the order of $R_E = 1.4 \cdot 10^{-8} (M/M_\odot)^{1/2} (D/kpc)^{1/2}$, which corresponds to impact parameters of the order of $1AU$. This is a completely different observational situation compared with experiments in the Solar system, where the best information regarding the propagation of light and radio signals is obtained from observations with impact parameters of the order of several $R_\odot$. The relative error of the existing experiments is of the order of $\sim 10^{-5}$ in view of estimates of the PPN parameter $\gamma$ [1]. However, if recalculated to distances of about $1AU$, there would be loss of accuracy by two orders of magnitude in the relative error. Nevertheless, this is significantly more accurate than our estimate.

This raises the question, how to improve the resulting estimate based on microlensing events? Obvious way is to include more data in the treatment. In this paper we reviewed the first 700 events from data for 2018 to collect 100 events in the sample. Not speaking about the future observations, we note that the Phase IV of the OGLE project [14] contains the data on about 16 thousand microlensing events (the previous phases are much smaller). This presumably allows one to increase the sample by $\approx 23$ times and, accordingly, to reduce $\sigma_{\langle\varepsilon\rangle}$ by about five times. This indicates good prospects for using the method.

## ACKNOWLEDGEMENT

This research essentially uses the data compiled by the Optical Gravitational Lensing Experiment, Phase IV, 2018 (http://ogle.astrouw.edu.pl/ogle4/ews/ews.html) [13, 14].

## APPENDIX A.

Table 1A. The sample of microlensing light curves from the OGLE database (OGLE-2018-BLG-XXXX).

| XXXX | | | | | | | | | |
|------|------|------|------|------|------|------|------|------|------|
| 0003 | 0097 | 0182 | 0247 | 0304 | 0409 | 0448 | 0541 | 0596 | 0655 |
| 0009 | 0098 | 0187 | 0249 | 0305 | 0410 | 0493 | 0545 | 0612 | 0658 |
| 0012 | 0100 | 0193 | 0261 | 0315 | 0411 | 0506 | 0566 | 0613 | 0660 |
| 0028 | 0105 | 0197 | 0263 | 0317 | 0418 | 0512 | 0567 | 0615 | 0662 |
| 0031 | 0127 | 0198 | 0287 | 0339 | 0419 | 0514 | 0574 | 0618 | 0663 |
| 0034 | 0137 | 0199 | 0288 | 0346 | 0421 | 0515 | 0575 | 0622 | 0672 |
| 0035 | 0142 | 0205 | 0292 | 0367 | 0423 | 0516 | 0576 | 0629 | 0681 |

| 0038 | 0149 | 0216 | 0293 | 0371 | 0436 | 0518 | 0579 | 0633 | 0711 |
| ---- | ---- | ---- | ---- | ---- | ---- | ---- | ---- | ---- | ---- |
| 0075 | 0156 | 0232 | 0300 | 0393 | 0437 | 0536 | 0581 | 0639 | 0713 |
| 0077 | 0171 | 0236 | 0303 | 0403 | 0447 | 0538 | 0589 | 0644 | 0727 |

**REFERENCES**

1. C.M.Will. The Confrontation between General Relativity and Experiment. Living Reviews in Relativity. 2014. **17**, id 4.

2. Александров А.Н., Вавилова И.Б., Жданов В.И., Жук А.И., Кудря Ю.Н., Парновский С.Л., Федорова Е.В., Яцкив Я.С. Общая теория относительности: признание временем. К.: Наукова Думка, 2015.

3. Schneider P., Ehlers J., Falco E.E. Gravitational Lenses. Berlin: Springer, 1992.

4. Dark energy and dark matter in the Universe: in three volumes / Ed. V. Shulga. – Vol. 2. Dark matter: Astrophysical aspects of the problem / V.M. Shulga, V.I. Zhdanov, A.N. Alexandrov, P.P. Berczik, E.P. Pavlenko, Ya.V. Pavlenko, L.S. Pilyugin, and V.S. Tsvetkova. – K.: Akademperiodyka, 2014. –356 p. (Chapters 1 and 2.)

5. Paczynski B. Gravitational Microlensing by the Galactic Halo. ApJ. 1986. 304. 1.

6. Keeton C.R., Petters A.O. Formalism for testing theories of gravity using lensing by compact objects. I: Static, spherically symmetric case. Phys. Rev. 2005. D72. 104006.

7. Keeton C.R., Petters A.O. Formalism for testing theories of gravity using lensing by compact objects. II: Probing Post-Post-Newtonian metrics. Phys. Rev. 2006. D73. 044024.

8. Lubini M., Tortora C., Näf J., Jetzer P., Capozziello S. Probing the dark matter issue in f(R)-gravity via gravitational lensing. Eur. Phys. J. C. 2011. 71(12). 1834.

9. Milgrom M. Testing the MOND Paradigm of Modified Dynamics with Galaxy-Galaxy Gravitational Lensing. Phys. Rev. Lett. 2013. 111. 041105.

10. Fedorova E., Sliusar V.M., Zhdanov V.I., Alexandrov A.N., Del Popolo A., Surdej J. Gravitational microlensing as a probe for dark matter clumps. MNRAS. 2016. 457. P.4147–4159.

11. Liu H., Wang X, Li H., Ma Y. Distinguishing *f(R)* theories from general relativity by gravitational lensing effect. Eur. Phys. J. C. 2017. 77. 723.

12. Александров О.М., Жданов В.І., Слюсар В.М. Перевірка формули Ейнштейна для гравітаційного відхилення світла за кривими блиску мікролінзованих джерел. Вісн. Київ. нац. ун-ту. Астрономія. 2019. 59, в.1. С.18-22.

13. Udalski A., Szymanski M., Kaluzny J., Kubiak M., Krzeminski W., Mateo M., Preston G.W., Paczynski B. The Optical Gravitational Lensing Experiment. Discovery of the First Candidate Microlensing Event in the Direction of the Galactic Bulge. Acta Astron. 1993. 43. P. 289–294.

14. Udalski A., Szymański M.K., Szymański G. OGLE-IV: Fourth Phase of the Optical Gravitational Lensing Experiment. Acta Astron. 2015. 65. P. 1-38.

15. Alcock C., Allsman R. A., Alves D., Axelrod T. S., Baines D., Becker A. C., Bennett D. P., Bourke A., Brakel A., Cook K. H., et al. Binary Microlensing Events from the MACHO Project. ApJ. 2000. 541. 270.

16. Gaudi B.S.. Microlensing Surveys for Exoplanets. Annual Review of Astronomy and Astrophysics. 2012. 50. 411.

17. Tsapras Y., Cassan A., Ranc C., Bachelet E., Street R., Udalski A., Hundertmark M., Bozza V., Beaulieu J. P., Marquette J. B., et al. An analysis of binary microlensing event OGLE-2015-BLG-0060. MNRAS. 2019. 487. 4603.

18. Smith M.C., Mao Sh., Paczyński B. Acceleration and parallax effects in gravitational microlensing. MNRAS. 2003. 339, 925.

19. Poindexter Sh, Afonso C., Bennett D.P., Glicenstein J.-F., Gould A., Szymański M.K., Udalski A. Systematic Analysis of 22 Microlensing Parallax Candidates. ApJ. 2005. 633. 914.